\documentclass{eas}
\usepackage{graphicx}
\TitreGlobal{The Universe of Dwarf Galaxies}
\begin{document}
\def\HI{H{\sc i} }
\def\HII{H{\sc ii} }
\def\Ha{{\rm H}\alpha }
\def\rg{\rho_{\rm g} }
\def\Msun{M_{\odot} }
\def\Mpc2{\Msun/pc^2}
\def\tff{\tau_{\rm ff} }
\def\tSF{\tau_{\rm SF} } 
\def\Zsun{Z$_{\odot}$ }
\def\OH{{12\-+{\it log}(O/H)} }
\def\lNO{{\it log}(N/O) }
\def\cd{{\it chemo-dynamical} }

\title{The Morphological Origin of Dwarf Galaxies} 
\runningtitle{G. Hensler: The Morphological Origin of DGs}
\author{Gerhard Hensler}
\address{
University of Vienna, Institute of Astronomy, Tuerkenschanzst. 17, 1180 Vienna, Austria; \\
 \email{gerhard.hensler@univie.ac.at}
}
\begin{abstract}
Dwarf galaxies (DGs) serve as extremely challenging objects in extragalactic astrophysics.
Their origin is expected to be set as the first units in CDM cosmology.
Nevertheless they are the galaxy type most sensitive to environmental  influences and their division into multiple types with various
properties have invoked the picture of their variant morphological
transformations. Detailed observations reveal characteristics
which allow to deduce the evolutionary paths and to witness how the
environment has affected the evolution. Here we refer to general 
morphological DG types and review some general processes, most of
which deplete gas-rich irregular DGs. Moreover, the variety
of pecularities is briefly refered, but cannot be comprehensively
analyzed because of limited paper space.
\end{abstract}
\maketitle

\section{Introduction}

After a former conference talk 
on evolutionary models of dwarf galaxies (DGs), in particular, 
on dwarf irregular galaxies (dIrrs), the blue compact subclass 
(BCDs) and on starburst (SB)DGs, a senior colleague 
well-known as one of the pioneers of DG studies
asked me with a serious surprise, why
DG evolution is so complex and not a straight-forward scenario
although their structural appearance looks pretty simple. 
Nowadays this first impression of DGs has changed totally 
with the advent of large telescopes, new techniques, 
and of the accessible panchromatic view.
Within the last two decades detailed analyses of DGs has extended
but also stirred our view of the classical morphological DG types,
dIrrs, SBDGs, and dEs including satellite galaxies,
so-called dwarf spheroidals (dSphs), as their extension to lower 
brightness of about -5$^m$. 

In their studies of Virgo cluster DGs already \cite{sb84} found
that elliptical DGs (dEs) dominate the cluster galaxy population
by far. This stands in contrast to their number fraction in the 
field where dIrrs are the most frequent DGs but with a smaller 
fraction than cluster dEs (\cite{bin88}).
For the interpretation of this issue and the occurrence of 
enhanced SF in dIrrs \cite{sb84} emphasized already the necessity 
of various links between the DG types by morphological transitions.

From the $\Lambda$CDM cosmology the baryonic matter should settle
within Dark Matter (DM) halos which are originally prefered
to form low-mass units, so-called subhalos, and hierarchically 
accumulating to massive galaxies. 
If the baryonic matter would follow this bottom-up structure
formation, the subhalos should also assemble their gas at first
and by this also evolve with star fomation (SF) to become 
the oldest galactic objects in the universe. That this
picture seems to be too naive is simply understandable by three 
major physical principles: the first one is, that the gas assembly 
works on the free-fall timescale $\tff$, 
namely, dependent on the gas density as $\rg^{-1/2}$, 
because gas is accreted through gravitation. At second, 
the SF timescale $\tSF$ is per definition proportional 
to M$_g/\Psi$ with $\Psi$ as the SF rate that, on the other hand,
in the self-regulated SF mode depends on $\rg^2$ (\cite{koe95}).
Since lower galaxy masses lead to less dense gas, SF is stretched 
over time for low-mass galaxies. And at least, as SF couples 
to stellar energy release and since the counteracting
cooling process depends on $\rg^2$, the gas expands due to 
pressure support and reduces the SF rate so that the effect 
of SF self-regulation is non-linearly amplified.

Another important effect that seems to affect the whole network
of galaxy formation and evolution is ionizing radiation from the
first cosmic objects to the re-ionization of the gas in the universe.
This changed not only its thermodynamical state so that its accretion 
onto low-mass objects was reduced (\cite{dij04}) but also evaporated gas 
that was already caught in minihalos (\cite{bl99}). 
Since massive objects remained
almost unaffected by the re-ionization phase, while DGs should have
experienced delayed SF (\cite{noe07}), this evolutionary dichotomy 
is observed as downsizing (\cite{cow96}). 
Nevertheless, the assumption that all DGs were affected in the 
re-ionization era and in the same way would request overlapping 
Stroemgren bubbles in an almost uniformly ionized universe. 
This assumption, however, is questioned and contrasted by the 
existence and amplification of density structures (\cite{par10}).

\section{Dwarf irregular and dwarf starburst galaxies}
\label{dIrrs}

dIrrs are characterized by large gas fractions, ongoing SF, 
and low metallicities $Z$. That dIrrs consist of the same 
or a higher gas fraction than giant spiral galaxies 
and mostly suffer the same SF efficiency, but appear with a 
wide range, but lower $Z$ than spirals, cannot be 
explained by simple evolutionary models. 
When gas is consumed by astration but replenished partly by 
metal-enriched stellar mass loss, the general analytical 
derivation relates the element enhancement with the 
logarithm of the decreasing remaining gas content where the 
slope is determined by the stellar yields 
(see e.g.\ textbooks like by \cite{pag10} or reviews as e.g. 
by \cite{hen10}). 
The effective yields of gas-rich galaxies decrease, however, with 
smaller galaxy masses (\cite{gar02,zee01}). This means that their
element abundances, particularly O measured in \HII regions, 
are much smaller than those released by a stellar population 
and confined to a "closed box". 

Two processes can reduce the metal abundances 
in the presence of old stellar populations: 
loss of metal-enriched gas by galactic outflows or infall 
of metal-poor to even pristine intergalactic gas (IGM).
It is widely believed, that a fundamental role in the 
chemical evolution of dIrrs is played by galactic winds,
because freshly produced metals in energetic events are
carried out from a shallow potential well of DGs through a 
wind (which will be therefore metal-enhanced).
Some SBDGs are in fact characterized by galactic winds 
(\cite{mar95}) or by large expanding supernova type II 
(SNeII)-driven X-ray plumes (e.g. \cite{hen98,mar02}). 
Studies have raised doubts to whether the expanding 
$\Ha$ loops, arcs, and shells mostly engulfing the X-ray plumes, 
lead really to gas expulsion from the galaxies 
because their velocities are mostly close to escape, 
but adiabatic expansion against external gas tends to 
hamper this. 

As an extreme, \cite{bab92} speculated that galactic winds 
are able to empty DGs from its fuel for subsequent SF events 
and, by this, transform a gas-rich dIrr to a fading gas-poor 
system.
In order to manifest this scenario and to study mass and
element abundance losses through galactic winds numerous 
numerical models are performed under various, but mostly 
uncertain conditions and with several simplifications 
(e.g. \cite{mlf99,str04}). 
The frequently cited set of models by MacLow \& Ferrara 
(self-gravitating, rotationally supported, isothermal \HI disks 
of dIrrs with fixed structural relations for four different gas 
masses between M$_{\rm g} = 10^6 - 10^9 \Msun$ and three 
different SNII luminosities in the center corresponding to 
SN rates of one per $3\times 10^4$ yrs to 3 Myrs) is mostly 
misinterpreted:  The hot gas is extremely collimated from the
center along the polar axis, but cannot sweep-up sufficient 
surrounding ISM to produce significant galactic mass loss.
On the other hand, the loss of freshly released elements from
massive stars is extremely high. Moreover, these models
lack of realistic physical conditions, as e.g. the existence of
an external pressure, self-consistent SF rates, a multi-phase
ISM, etc.

\begin{figure}[h]
\vspace{-1.0cm}
\begin{tabular}{ll}
\resizebox{0.48\columnwidth}{!}{%
  \includegraphics{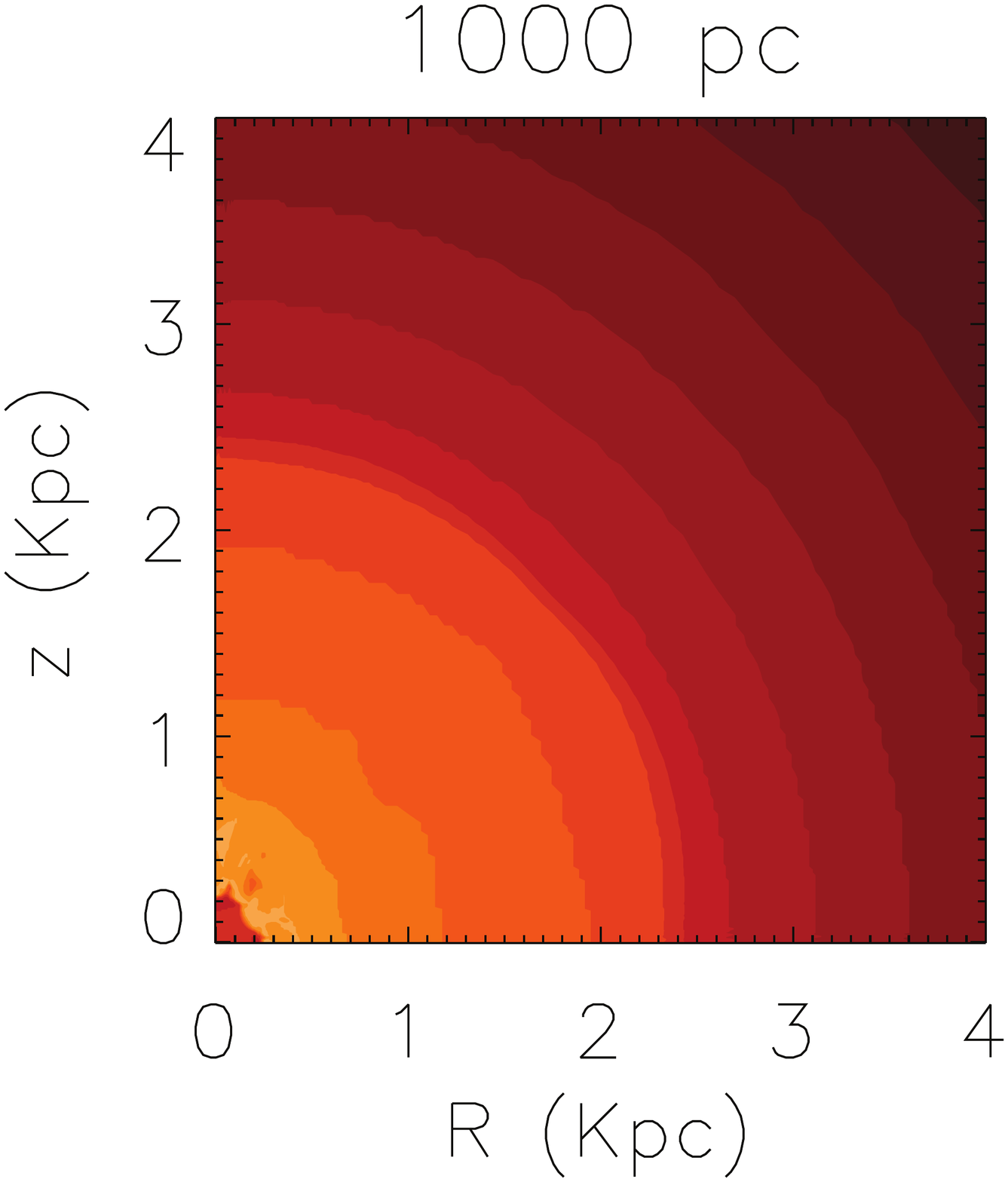} }
&
\resizebox{0.48\columnwidth}{!}{%
  \includegraphics{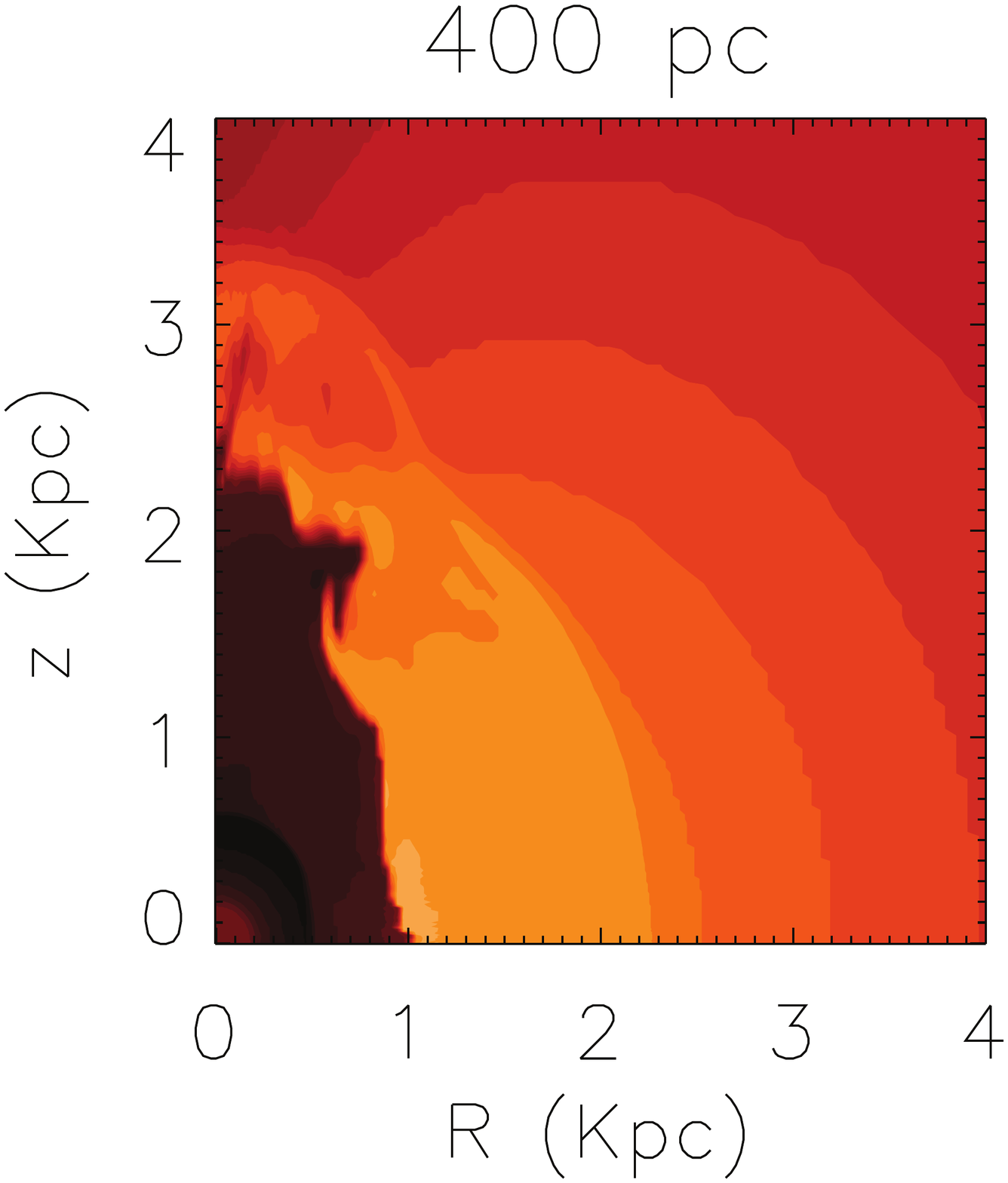} }
\end{tabular}
\vspace{-1.0cm}
\caption{
Density contours after 200 Myr of evolution for models differing on 
the semi-minor axis b of their initial configurations (semi-minor axis indicated on top of each panel),
while the major axis is set to a=1 kpc:  
{\it left:} b=a=1 kpc and spherical symmetric; 
{\it right:} b=400 pc, i.e. a disk eccentricity amounts 
to (a-b)/a=0.6. 
The density scale reaches from 10$^{-27}$ (black) to 10$^{-24} g/cm^3$ (brightest). {\it (for details see \cite{rec09}.)} 
}
\label{fig:wind}
\end{figure}

Also more detailed numerical simulations (\cite{db99}; \cite{rec06a}), 
show that galactic winds are not very effective in removing gas 
from a galaxy. Although galactic winds develop vertically, 
while the horizontal transport along the disk is very limited,
their efficiency depends very sensibly on the galaxy structure 
and ISM properties, as e.g. on the \HI disk shape (\cite{rec09}).
Fig. \ref{fig:wind} reveals clearly that the more eccentric the disk 
is, the more pronounced does the superbubble expand.
On the one hand, the hot SN gas has to 
act against the galactic ISM, exciting turbulence and mixing
between the metal-rich hot with the surrounding \HI gas. Not
taken into account in present-day models is the porosity
of the ISM, consisting of clouds and diffuse less dense gas.
In particular, the presence of clouds can hamper the development 
of galactic winds through their evaporation. 
This so-called mass loading reduces the wind momentum and internal 
energy. Since the metallicity in those clouds are presumably lower
than the hot SNII gas, also the abundances in the outflow are 
diminished as e.g. observed in the galactic X-ray outflow of
NGC~1569 (\cite{mar02}) for which a mass-loading factor of 10
is derived to reduce the metallicity to 1-2 times solar.
In recent simulations \cite{sca10} demonstrate that turbulent 
mixing can effectively drive a galactic wind. Although they
stated that their models lead to a complex, chaotic distribution 
of bubbles, loops and filaments as observed in NGC~1569, 
other observational facts have not been compared.

Detailed numerical simulations of the chemical evolution 
of these SBDG by \cite{rec06b}, e.g., could simultaneously 
reproduce both, the oxygen abundance in the warm gas as well as 
the metallicity in the hot outflow. \cite{rec07} show that 
the leakage of metals from a SBDG is surprisingly not 
prevented by the presence of clouds, because they
pierce holes into the wind shells. This leads to a final 
metallicity of only a few tenths of dex lower than in models 
without clouds.
 
Consequently, the basic question must be answered which physical 
processes trigger such enormous SF rates as observed 
in SBDGs and would consume all the gas content within much less 
than the Hubble time. One possibility which has been favoured
until almost two decades ago was that at least some of these
objects are forming stars nowadays for the very first time.
Today it is evident that even the most metal-poor ones 
(like I~Zw~18) contain stars of at least 1 Gyr old (\cite{mom05}),
but most SBDGs have several Gyrs old stellar populations.
This means that SF in the past should have proceeded in dIrrs, 
albeit at a low intensity what can at best explain their 
chemical characteristics, like for instance the low 
[$\alpha$/Fe] ratio.

\begin{figure}[h]
\vspace{-0.5cm}
\begin{tabular}{ll}
\resizebox{0.6\columnwidth}{!}{%
  \includegraphics[width=7cm]{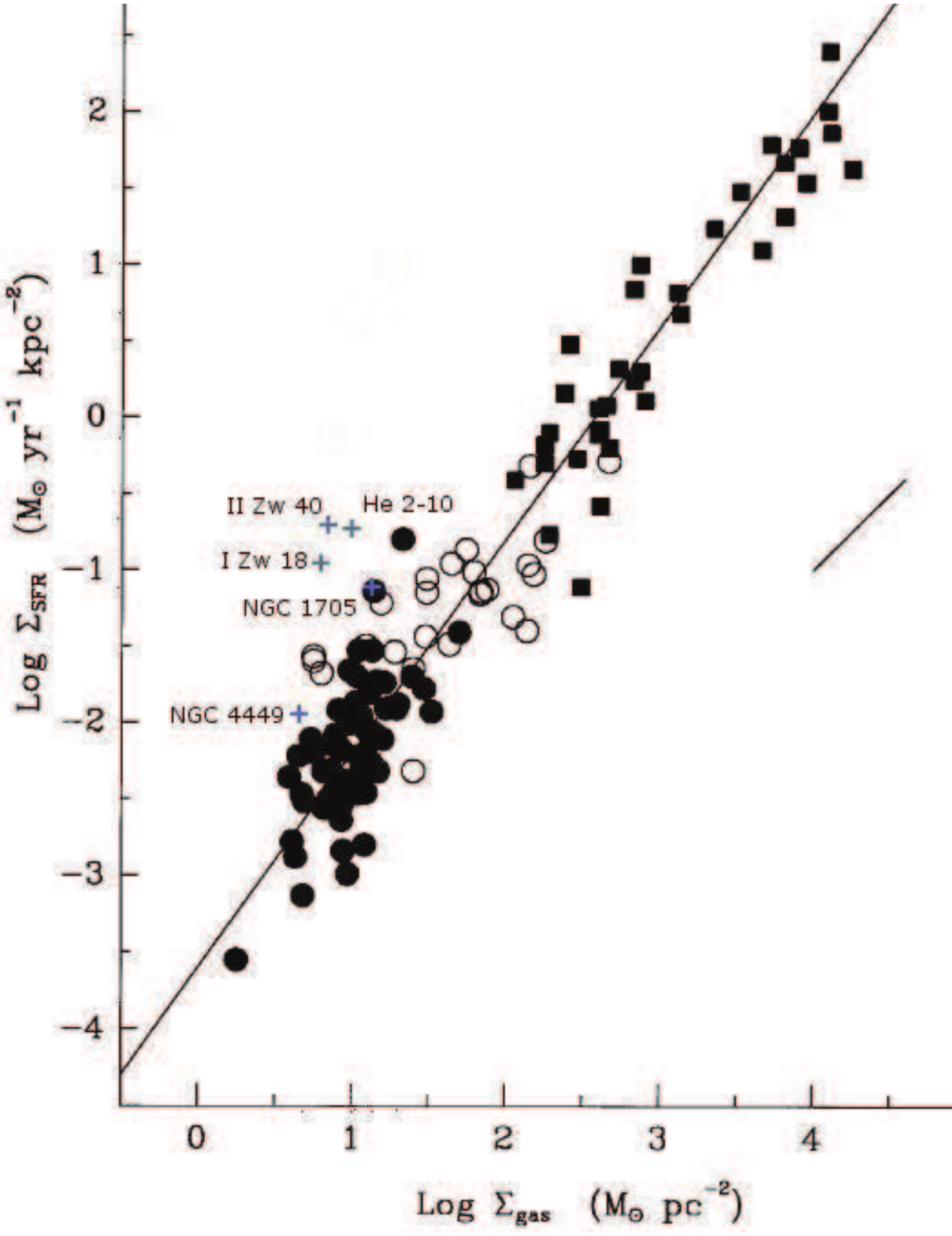} }
&
\begin{minipage}[t]{4.0cm}
\vspace{-9.0cm}
\caption{
 Comparison of the star-formation rates vs. gas content, in the
 optical galactic body of a few prototypical starburst dwarf 
 galaxies denoted by crosses and names with the well-known
 Kennicutt-Schmidt relation derived by \cite{ken98} with 
 an exponent of 1.4 (long full-drawn line).  
 {\it (from \cite{kue11})}  
} 
\end{minipage}
\hfill
\end{tabular}
\vspace{-0.5cm}
\label{fig:KS} 
\end{figure}

In most SBDGs large \HI reservoirs, however, enveloping the 
luminous galactic body have been detected 
(NGC~1705: \cite{meu98}, 
NGC~4449: \cite{hun98}, 
NGC~4569: \cite{sti02},
NGC~5253: \cite{kob08}, 
I~Zw~18: \cite{zee98c}, 
II~Zw~40: \cite{zee98b})
with clearly disturbed gas kinematics and disjunct from the 
luminous galactic body. Nevertheless, in not more than
two objects, NGC~4569 (\cite{mue05}) and NGC~5253 (\cite{kob08})
gas infall is proven, while for the other cases the gas kinematics
obtrudes that the gas reservoir feeds the engulfed DGs. 
In another object, He~2-10, the direct collision of an intergalactic
gas cloud with a DG (\cite{kob95}) is obviously triggering a huge SB.

Yet it is not clear, what happens to dIrrs if they experience on
an increasing external pressure what would happen e.g. when they 
fall into galaxy clusters. In sect.\ref{dEs} we will discuss the
effect of ram pressure on the structure of the ISM for which
numerical models exist for spiral galaxies (e.g. \cite{roe05})
as well as for dIrrs (e.g. \cite{mor00}), but only hints from 
observations. The effect on the SF rate due to compression 
of the ISM is observed, but not yet fully understood. 
\cite{cor06} e.g. observed a coherent enhancement of SF in 
group galaxies falling into a cluster, therefore, denoted 
as blue infalling group. 
 
The [$\alpha$/Fe] vs. [Fe/H] behaviour is representative of
the different production phases, $\alpha$-elements from 
the short-living massive stars and iron to 2/3 from type Ia SNe 
of longer-living binary systems. If the SF duration in a galaxy 
is very short, type Ia SNe do not have sufficient time to 
enhance the ISM with Fe and most of the stars will be overabundant 
in [$\alpha$/Fe]. The low average [$\alpha$/Fe] ratios in 
dIrrs, however, compared to large galaxies serve as a hint
of a long-lasting mild SF in these galaxies (\cite{lm04}).

\begin{figure}[h]
\vspace{-0.5cm}
  \includegraphics[width=12cm]{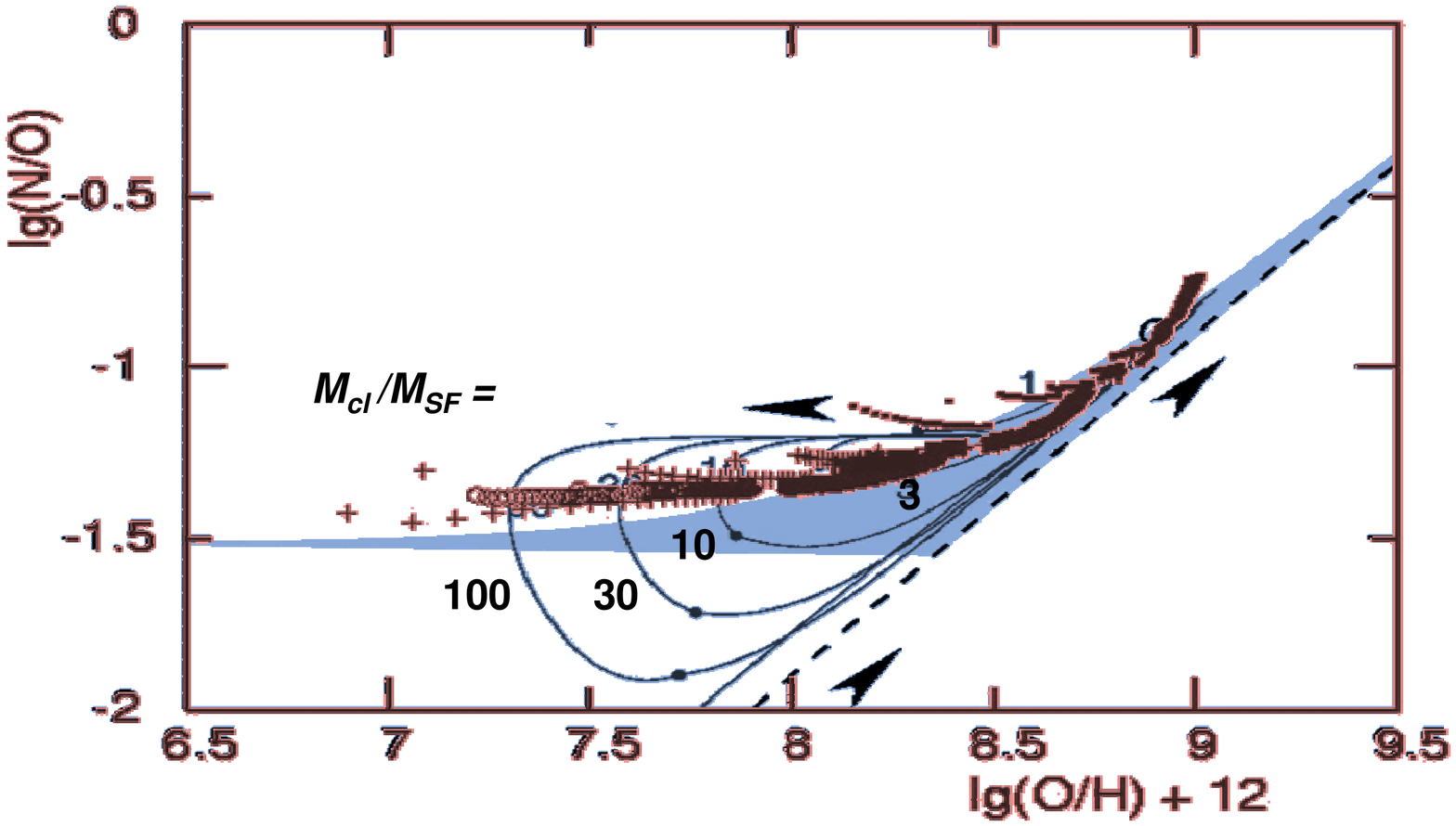} 
\vspace{-2.5cm}
\caption{
The abundance ratio N/O as a function of oxygen abundance
observed in spiral and irregular galaxies (shaded area;
after: \cite{zee98a}) overlayed with evolutionary loops due to 
infall of primordial intergalactic gas clouds. 
These have different mass fractions $M_{cl}/M_{SF}$ with 
respect to the mass involved in the SF region The crosses
represent evolutionary timesteps of models, the arrows depict
the direction of the evolutionary paths.
The dashed straight line represents a simple model relation  
for purely secondary nitrogen production. For discussion see: 
\cite{koe05}). }
\label{fig:NO} 
\end{figure}

Although the mass-metallicity relation also holds for dIrrs
and even steepens its slope (\cite{tri04}) what can be interpreted 
by galactic mass loss and the corresponding lower effective
yield (\cite{gar02}), the abundance ratios are unusual. As
mentioned above O/Fe reaches already solar values for 
subsolar oxygen abundances. While this can be explained
by means of a long SF timescale, another characteristic and 
unusual signature is that with O abundances below 1/10 solar
the N/O ratio \lNO remains at about -1.6, i.e. clearly 
smaller than in gSs and with a large scatter but no significant 
correlation with oxygen (see Fig.\ref{fig:NO}). 
Their regime of N/O--O/H values overlaps with those of 
\HII regions in the outermost disk parts of gSs at around 
\OH = 8.0 ... 8.5 (\cite{zee98a}). 

In the 90th several authors
have tried to model these observations by SF variations with
gas loss through galactic winds under the assumptions that these
dIrrs and BCDGs are young and experience their first epochs of 
SF (for a detailed review see \cite{hen99}). Since stellar 
population studies contradict to the youth hypothesis, another 
process must be invoked. Since these objects are embedded into 
\HI envelops and are suggested to suffer gas infall as manifested
e.g. for NGC~1569 (see above, \cite{sti02,mue05}), the influence 
of metal-poor gas infall into an old galaxy with continuous SF 
on particular abundance patterns should be exploited. 
With the reasonable approach that the fraction of infalling gas 
increases with decreasing galaxy mass, 
their results could match not only the observational regime of 
BCDs in the [\OH]-\lNO space but also explain the shark-fin shape 
of observational data distribution (\cite{koe05}).

\section{Dwarf elliptical galaxies}
\label{dEs}

dEs are the most numerous type in clusters and are frequently 
denoted as examples of ``stellar fossil'' systems in which 
the bulk of their SF occurred in the past.  
They are preferentially located in morphologically evolved 
environments (\cite{tre02}), i.e. in regions with high 
galaxy densities and dominate the morphological types of galaxy 
in galaxy clusters, as e.g. Virgo, Coma, Fornax, and Perseus. 
Furthermore, \cite{tul08} observed that dEs strongly cluster around 
luminous elliptical/S0 galaxies. The evolution of this galaxy type
 should be mainly caused by gas and tidal effects on SF and 
structure and indicates that it is strongly affected by environment.

Already \cite{bot85} found that cluster dEs are usually almost 
free of interstellar gas and contain few young stars.  
In trying to understand the dE population, structural regularities 
and correlations must be studied, as it is known since the 80th 
between optical surface brightness and luminosity (\cite{bin84,kor85}) 
and between luminosity and stellar velocity dispersion  
which also correlate with metallicity (e.g. \cite{pet93}).
Furthermore, dEs often have flattened profiles but are mostly
kinematically supported by their stellar velocity dispersions rather 
than by rotation (\cite{ben91}). 

The combination of low gas-mass fractions and moderate-to-low stellar 
metallicities in dE (about 0.1 of solar or less) is a key feature of 
this class. 
The lower abundances of stars in dEs (\cite{haa97}) suggest that 
extensive gas loss occurred during their evolution and SF ceased 
due to a lack of raw material rather than exhaustion of the gas 
supply through SF. 
Galactic winds are therefore a hallmark of modern models for dE 
galaxies, starting from the basic consideration by \cite{lar74} 
and continued with the study by \cite{ds86}. They are commonly 
assumed to have cleaned out dE galaxies soon after their 
formation. As mentioned in sect. \ref{dIrrs}  gas
expulsion by means of galactic winds even in low-mass systems 
requires a dark-to-baryonic matter ratio (\cite{mlf99}) much 
smaller than assigned to DGs in the classical formation picture 
(e.g. \cite{mat98}). There are two competing scenarios for the 
formation of dEs. On the one hand, those low-mass galaxies are 
believed to constitute the building blocks in $\Lambda$CDM 
cosmology and should therefore have evolved congruently with 
the mass accumulation to the much more massive entities, galaxies
and galaxy clusters, by this leading to SF with the downsizing 
effect through the delay by the re-ionization epoch. Their stellar
component is expected to be heated continuously by harrasment 
of more massive cluster galaxies and thus to be pressure supported.

A variety of observations are available which also support
diverse scenarios of dEs evolution (see e.g. review by 
Lisker, conference).   
Recent HI studies of Virgo cluster dEs (\cite{con03}) 
and also those of the Fornax cluster (e.g.\ \cite{mic04})
have unveiled that a small but significant fraction of them 
contains gas, has experienced recent SF, and can be argued 
from internal kinematics and cluster distribution data to represent 
an infalling class of different types of gas-rich galaxies in the 
state of morphological transformation. Recent findings of a 
significant fraction of rotationally supported dEs in the 
Virgo cluster (\cite{zee04}) also supports the possibility 
of morphological transformation from dIrrs to dEs thru gas 
exhaustion (\cite{bos08}). This separation should therefore 
also be visible in an intermediate age stellar population, 
flatter figure shape, and rotation.
Indeed, \cite{lis07} found that dEs in the Virgo cluster can be 
divided in different subclasses which differ significantly 
in their morphology and clustering properties, however do not 
show any central clustering, but are distributed more like 
the late-type galaxies. 
These subclasses show different disk signatures, such as 
bars and spiral structures. Also these types of dEs are not
spheroidal, but rather thick disk-like galaxies.  

Similar shapes were also found for the brighter, non-nucleated dEs.
There is only a small fraction of nucleated dEs whithout any 
disk features or cores, which keep the image of spheroidal objects 
consisting of old stars.

\begin{figure}[h]
\vspace{-1.0cm}
\begin{tabular}{ll}
\resizebox{0.6\columnwidth}{!}{%
  \includegraphics[width=7cm]{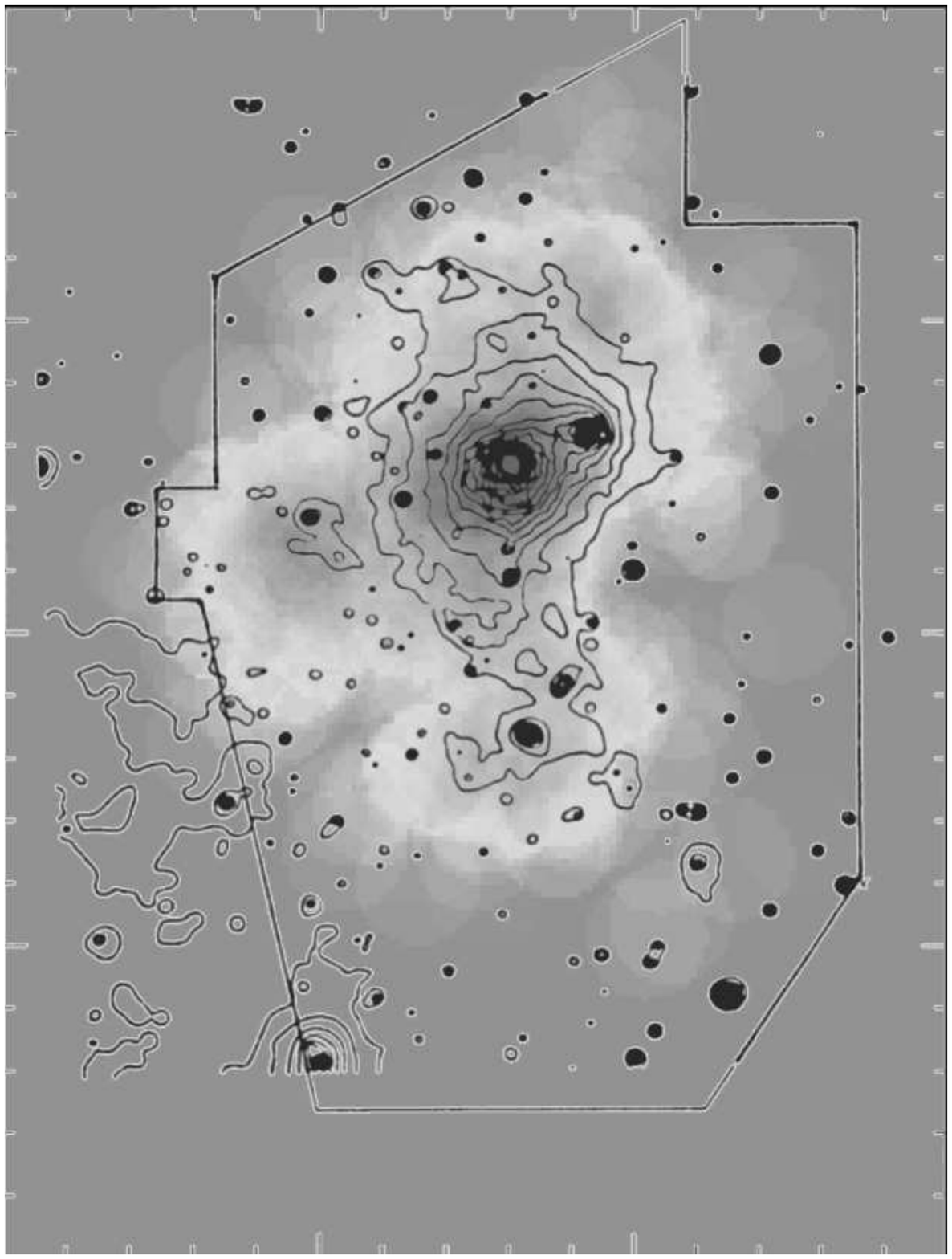} }
&
\begin{minipage}[t]{4.0cm}
\vspace{-9.0cm}
\caption{
Distribution of dEs in the Virgo Cluster divided into 
rapid (white) and slow objects (darker central part) 
overlaid with X-Ray brightness contours 
{\it (courtesy by Thorsten Lisker)}.
}
\end{minipage}
\hfill
\end{tabular}
\vspace{-1.0cm}
\label{fig:VCdE} 
\end{figure}

A figure analysis of Virgo dEs correlates with the averaged 
orbit velocity in the sense that flatter (infalling and transformed) 
dEs show on average a larger orbital velocity 
(700 km/s) than those originating within the cluster (300 km/s)
(\cite{lis09}). This kinematical dichotomy is expected 
because galaxies formed in virial equilibrium with the cluster
retain their initial kinetic energy while the cluster mass grows. 
Galaxies falling into the present cluster potential must 
therefore obtain larger velocities. 

To obtain information about their both evolutionary stages, 
the young infalling vs. the late cluster members, \cite{got09} 
studied SDSS data. The basic model is that dIrrs which are formed 
outside the Virgo Cluster and becoming stripped on their infall,
by this being transformed into dEs, should reveal properties 
recognizably different from dEs which have already aged in the 
cluster, as e.g. colors, effective radius, radial stellar 
distribution, and abundances. One result by \cite{got09} is that 
for the two dE populations, with and without cores, distinguished 
by their Sersic parameter, there is only a slight indication 
that non-nucleated dEs are more concentrated towards the inner 
cluster regions, whereas the fraction of nuclated dE is
randomly distributed, while \cite{lis07} found it to increase
with distance. 
An analysis of the relation between the central surface brightness
and the Sersic parameter shows the expected tendency to higher values 
for brighter galaxies. Furthermore, there were no relations found 
for the Sersic parameter, the effective radius, or the 
distance from M87.

For deeper insights spectra are urgent. And really, \cite{tolo09}
derived for Coma cluster dEs to be weaker in carbon than dEs 
in low-density environments, while they are similar in nitrogen.
Most recently, \cite{tol10} found that pressure supported Virgo dEs 
show higher dynamical mass-to-light ratios than rotationally 
supported dEs of similar luminosity and further that dEs 
in the outer parts of the cluster are mostly rotationally 
supported with disky shapes. 

Rotationally supported dEs even follow the Tully-Fisher relation. 
One of \cite{tol10} fundamental and most spectacular results is, 
however, that dEs are not DM-dominated galaxies, at least 
out to the half-light radius. Concerning any metallicity
gradient, the picture is not yet clear, but also not what one
should expect. While \cite{spo09} show a tight positive correlation 
between the total metallicity $Z$ and the mass, \cite{kol09} 
do not find any trend involving [Fe/H] for Fornax-cluster and 
nearby-group dEs. However, they found metallicity gradients to 
exist and argue that this is a lack of sufficient mixing of 
old stellar populations but compare it also with simulations 
which found this as a result of galactic winds.  
Moreover, from the deconvolution of the SF history of their 
sample dEs with respect to the central 1 arcmin and within 
the effective radius \cite{tolo09} allow the conclusion on
the existence of SF episodes in the very center even within 
the last 1 Gyr for a few objects.
Correlations of both signatures, SF history and metallicity
gradients, for cluster-member dEs vs. infall dEs should be
derived for more clusters, but observations are unfortunately 
very time-expensive if possible at all.

\section{Dwarf spheroidal galaxies}
\label{dSphs}

At the faint end of dEs another type of almost gas-free spheroidal 
DGs exists which is located around massive galaxies like our 
MWG and M31 and thus orbiting them as satellites.
These objects have attracted increasing attention over the last years 
because at their low-mass end they overlap with Globular Clusters, 
so that the understanding of their formation and evolution is of 
substantial relevance for our astrophysical picture of cosmology 
and galaxy evolution. Four main questions are addressed: 
\vspace{-0.3cm}
\begin{enumerate}
\item How and when did they form? They all harbour a very old stellar 
population (\cite{tols09}) and seem to have evolved unaffected of 
the re-ionization era (\cite{gre04}). \vspace{-0.3cm}
\item Is their existence as satellite system typical for all massive galaxies? Their origin and DM content is still questioned 
by some authors (\cite{kro10}). The large discrepancy of
the number of objects really observed and the one expected from
$\Lambda$CDM cosmology and because of their orbit concentration 
to the so-called disk of satellites, also observed around M31, 
invoked the preferrence of their tidal-tail origin (\cite{met09}).
The large velocity disperion is caused by the tidal effects. 
\vspace{-0.3cm}
\item How is their evolution determined by the vicinity of the 
massive mature galaxy? Not only the tidal field must have a 
disruptive effect, but also a gaseous halo of the central galaxy 
will interact with the ISM of the dSphs (\cite{may07}).
The tendency of an increasing gas fraction bound to the dSphs 
with distance from the MWG (\cite{har01}), points into that 
direction. \vspace{-0.3cm}
\item Vice versa the question arises, how the satellites influence 
the structure and evolution of the mature galaxy, here the MWG.
\end{enumerate}
 \vspace{-0.2cm}

The first three questions also concern the morphological origin 
of dSphs e.g. as tidal-tail DGs formed in the same era as the MWG 
but in a gas-rich galaxy collision or their transition from 
gas-rich $\Lambda$CDM satellites to dSphs. 

The fascinating wealth of data and their precision on stellar ages
and kinematics, on their chemical abundances, abundance gradients, 
and tidal tails of dSphs 
(for most recent reviews see e.g. \cite{koc09} and \cite{tols09}) 
have triggered numerous numerical models which, however, still lack 
of the inclusion of the above-mentioned environmental effects. 
Not before recently \cite{ph11}; first insights also in this 
conference proceedings and in \cite{hen11}) have started 
to simulate the full sample of a large number of DM subhalos 
containing baryonic matter by \cd models. Since these models 
cannot be presented here because of paper space, we wish to 
highlight only the main issues as follows and refer the 
interested reader to a forth-coming paper by \cite{ph11}).
In addition to the tidal and gaseous effects by means of the MWG, 
due to the mutual interactions also between the dSphs disruption, 
gas and star exhaustion, and merging occur. The metallicity and
abundance ratios are traced and comparable with observation. 
The SF histoies vary dependent on mass and gas re-accretion.

\section*{Acknowledgement}
\vspace{-0.3cm}
The author is grateful to Joachim Koeppen, Thorsten Lisker,
Mykola Petrov, Simone Recchi, Werner Zeilinger, and the
SMAKCED collaborations for enlightening and continuous 
discussions on DGs.

%

\end{document}